\documentclass[8pt,aps,twocolumn]{revtex4}
\usepackage{amssymb}
\usepackage{latexsym}

\usepackage{epsfig}

\usepackage{dcolumn}
\usepackage{amsmath}
\usepackage{amsfonts}
\usepackage{bm}

\newcommand{\be}{\begin{equation}}
\newcommand{\ee}{\end{equation}}
\newcommand{\bea}{\begin{eqnarray}}
\newcommand{\eea}{\end{eqnarray}}

\newcommand{\p}{\partial}
\newcommand{\s}{\sigma}

\newcommand{\la}{\langle}
\newcommand{\ra}{\rangle}
\newcommand{\rd}{\mbox{d}}
\newcommand{\ri}{\mbox{i}}
\newcommand{\re}{\mbox{e}}

\begin{document}
\title{Zero energy Majorana modes in superconducting wires.}
\author{A.M. Tsvelik}
\affiliation{Department of Condensed Matter Physics and Materials Science, Brookhaven National Laboratory,
  Upton, NY 11973-5000, USA} \date{\today } \begin{abstract} I consider a model of two thin superconducting wires interacting through Josephson tunneling and Coulomb interactions. It is shown that there is a parameter range where such a model may have two ground states with different types of quasi-long-range order: one is superconducting and the other is pair density wave. It is further shown that boundaries between these different phases support zero energy Majorana modes and that these modes are robust against disorder.  \end{abstract}


\maketitle
%
%

Majorana fermions have attracted a great deal of attention lately. A part of this interest is related to the fact that zero energy Majorana modes (ZEMMs) are nonlocal and robust with respect to all sorts of perturbations. The latter property is related to the fact that ZEMMs emerge as  bound states of Majorana fermions and  topological objects such as vortices \cite{kopnin},\cite{read},\cite{ivanov} or kinks \cite{yakovenko}. The nonlocality is related to the fact that to define an occupation number operator one needs at least two ZEMMs. More formally, the dimensionality of the Hilbert space spanned by $N$ degenerate  Majorana modes is $2^{[N/2]}$. In particular,  to make a two-level system one needs at least two modes located at different positions.   Thus  spatially separated ZEMMs correspond to the situation described in the  Einstein-Podolsky-Rosen paradox. 

The Majorana fermion is not as exotic as it might seem, being a linear superposition of an electron and a hole, and hence a particular case of a Bogolyubov quasiparticle. Indeed, the most straightforward way of realizing Majorana fermions is to create them literally from electrons and holes and they have been suggested to emerge in p-wave superconductors \cite{ivanov},\cite{yakovenko}, topological insulators (see, for instance \cite{fu},\cite{roy}) or in semiconducting wires with strong spin-orbit coupling in contact with s-wave superconductors \cite{fisher}.  There are other possibilities, however. Majorana fermions may emerge as collective excitations as, for instance, in the one-dimensional transverse field Ising model or in the Kitaev model \cite{kitaev}. In these cases Majorana fermion operators are constructed from spin S=1/2 ones via a Jordan-Wigner transformation; the resulting objects are highly nonlocal in spins.

In the current paper I consider Majorana fermions in one-dimensional superconductors. However, instead of being Bogolyubov quasiparticles, they emerge as collective excitations; more precisely, as nonlocal objects made of superconducting pairs. Such a route may appear to be unnecessarily complicated, but, in fact, represents several advantages. One of these is that to get the desired result one does not need any exotic superconductors, like $p$-wave ones, but can manufacture Majorana fermions from conventional $s$-wave or high temperature $d$-wave superconductors. Here I am encouraged by the experimental observation of the magnetic field driven insulator-superconductor crossover in superconducting wires \cite{exp} and by its theoretical explanation given in \cite{shimshoni}.
 
Below I consider a model of two coupled superconducting wires. As is known from previous work \cite{shimshoni}, such a system can be in two different phases depending on the balance between the inter-wire Coulomb repulsion and Josephson tunneling energy. One phase supports superconducting quasi-long-range order (SC) and the other one supports pair density wave (PDW) quasi-long-range order. In the presence of disorder the PDW is pinned and this phase is an insulator. I demonstrate that boundaries between the SC and PDW phases support ZEMMs and this fact is not affected by disorder. I also show that in the state where domain walls between SC and PDW phases have a finite density a narrow Majorana band is formed with a density of states which has a sharp singularity of zero energy.


Consider two coupled superconducting wires separated by a thin  insulating layer of width $w$ as has been considered in  \cite{shimshoni},\cite{carr},\cite{tsvelik}. This construction can be also realized as a double chain Josephson junction array or, in the limiting case, as a system of two chains of attractive Hubbard models. Since all spin degrees of freedom are gapped, at low temperatures the dynamics of individual wires are described by the phase fields $\Phi_i$ ($i=1,2$). The wires are coupled by the Josephson pair tunneling and Coulomb interaction. 
\bea
&& H = H_1 + H_2 + V,\label{problem}\\
&& H_i = \frac{1}{2}\int \rd x\Big[U_0n_i^2 + \frac{\rho_s}{4m}(\p_x\Phi_i)^2\Big],\\
&& V = \int \rd x\Big[-g_J\cos(\Phi_1 - \Phi_2 - Qx) + U n_1n_2\Big], \label{V}
\eea
The operators $n_i$ denote density fluctuations of the Cooper pairs and can be represented as \cite{giam}:
\bea
n(x) = - \frac{1}{\pi}\p_x\Theta + n_0\sum_{p=-\infty}^{\infty} \re^{2\ri p(\pi n_0 x - \Theta)}\label{dens}
\eea
where $\Theta_i$ is the field dual to $\Phi_i$: $
[\p_x\Theta (x), \Phi(y)] = -\ri\pi\delta(x-y)$. 
As it was argued in \cite{shimshoni}, $Q$ is determined by the effective magnetic field and is given by the deviation of the vortex density from the nearest integer value $N$:
$Q = 2ewB/c -N$. Substituting (\ref{dens}) into (\ref{V}) and defining  the fields 
\bea
&& \Phi_{1,2} =  \sqrt\pi\Big(K_+^{1/2}\Phi_+ \pm K_-^{1/2}\Phi_-\Big), \nonumber\\
&&\Theta_{1,2} = \frac{\sqrt\pi}{2}\Big(K_+^{-1/2}\Theta_+ \pm K_-^{-1/2}\Theta_-\Big),\nonumber\\
&& [\p_x\Theta_a(x),\Phi_b(y)] = -\ri\delta_{ab}\delta(x-y),
\eea
I obtain the Hamiltonian $H = H_+ + H_-$, where $H_+$ describes  the symmetric mode $(+)$: 
\bea
H_+ = \frac{v_+}{2}\int \rd x \Big[(\p_x\Phi_+)^2 + (\p_x\Theta_+)^2\Big], \label{Gaussian}
\eea
and $H_-$ contains only the asymmetric fields: 
\bea
&& H_- = \int \rd x \Big\{ \frac{v_-}{2}\Big[(\p_x\Phi_-)^2 + (\p_x\Theta_-)^2\Big] + \label{charge2}\\
&& V_c\cos\Big(\sqrt{4\pi K_-}\Phi_- -Qx\Big)- V_J\cos\Big(\sqrt{4\pi/K_-}\Theta_-\Big)\Big\} \nonumber,
 \eea
 where $V_c = n_0^2U, V_J = n_0^2 g_J$ and 
 \bea
 K_{\pm} = \Big[\frac{m(U_0 \pm U)}{\pi^2\rho_s}\Big]^{1/2}, ~~ v_{\pm} = \Big[\frac{\rho_s(U_0\pm U)}{m}\Big]^{1/2}.
 \eea
 
 Depending on which of the cosines in (\ref{charge2}) takes over, the ground state of this model describes either quasi long range superconducting order or pair density wave.  The latter state has a singularity in the density-density correlation function at the finite wave vector $2\pi n_0$. 
 When both cosines are relevant (that is at $1/2 < K_- <2$) these states are separated by a quantum critical point (QCP), the location of which is approximately determined by the relation
 $(V_J/\Lambda)^{K_-} \sim (V_c/\Lambda)^{1/K_-}$, 
 where $\Lambda$ is the ultraviolet cut-off. Fulfillment of the above condition on $K_-$ is essential for the subsequent arguments. 
 
 The vicinity of the QCP can be studied analytically when $K _-\approx 1$ (this will be my assumption throughout the rest of the paper). In that case it is convenient to refermionize (\ref{charge2}) with the result 
  \bea
 &&  H_- =\nonumber\\
 && \int \rd x \Big\{ \frac{\ri v_-}{2}(-\rho_R\p_x\rho_R + \rho_L\p_x\rho_L - \eta_R\p_x\eta_R + \eta_L\p_x\eta_L) +\nonumber\\
 &&  4\pi v_-(K_- -1)\rho_R\rho_L\eta_R\eta_L +  2\ri m_+\rho_R\rho_L + 2\ri m_-\eta_R\eta_L\Big\}, 
\nonumber\\
&& m_{\pm} = V_J \pm V_c . \label{fin}
 \eea
 Here $\rho_{L,R}$ and $\eta_{L,R}$ are left- and right-moving components of Majorana (real) fermions. This model is equivalent to the continuum limit of two quantum Ising (QI) models coupled by the energy density operators.  It is useful to express important operators of the original problem (\ref{problem}) in terms of operators of the QI model. To this end I will use the standard correspondence between the bosonic exponents and the order $\s_{1,2}$ and disorder $\mu_{1,2}$ operators of the QI models about which all the relevant information can be found, for instance, in \cite{book}.  For instance, for the first chain 
the superconducting order parameter and the most relevant part of the oscillatory  density (\ref{dens}) are 
 \bea
 && \re^{\ri\Phi_1} = \\
 && \re^{\ri\sqrt{\pi K_+}\Phi_+}\re^{\ri{\sqrt\pi K_-}\Phi_-} = \re^{\ri{\sqrt\pi K_+}\Phi_+}\Big[\s_1\s_2 + \ri\mu_1\mu_2\Big].\nonumber\\
 && n_{osc}^{(1)}/n_0 = \re^{2\ri\Theta_1} =\label{stag1}\\
 && \re^{\ri\sqrt{\pi/K+}\Theta_+}\re^{\ri\sqrt{\pi/K_-}\Theta_-} = \re^{\ri\sqrt{\pi/K_+}\Theta_+}\Big[\s_1\mu_2 + \ri\mu_1\s_2\Big]. \nonumber
\eea
Since $\la\s\ra \neq 0$  when the corresponding fermionic mass is positive and zero when it is negative (for $\la\mu\ra$ it is the other way around), we see that in the phase where 
$m_+ = V_J +V_c >0, ~~m_- = V_J - V_c >0$ it is the superconducting order parameter which acquires a finite amplitude and in the phase $m_+ > 0, ~~m_- <0$ this happens for the oscillatory part of the density. 
 
 Neglecting the four-fermion interaction, I obtain from (\ref{fin}) that the amplitude of the superconducting order parameter diminishes when $V_J$ approaches $V_c$ from above and vanishes at $V_J \leq V_c$. Elementary scaling considerations yield the following estimate for the amplitude: 
 \be
 \la \s_1\s_2\ra \sim (V_J^2-V_c^2)^{1/8}.
 \ee
At $V_c > V_J$ one gets the same expression for the amplitude of the pair density wave. 

 The transition between two phases can be driven by a magnetic field. A magnetic field adds to (\ref{fin}) the term 
 \be
2\ri vQ(\eta_L\rho_L - \eta_R\rho_R).
\ee
(I drop the subscript ``minus" on the velocity). Then the spectrum becomes
\bea
&& \omega^2_{1,2} = (vk)^2 + (vQ)^2 + (V_J^2 + V_c^2) \pm \nonumber\\
&& 2\Big[(v^2k Q)^2 + v^2Q^2V_J^2 + V_c^2V_J^2\Big]^{1/2}.
\eea
The QCP where $\omega_-(k=0) =0$ is achieved for 
\be
(vQ)^2 = m_+m_- = (V_J^2-V_c^2).
\ee
The magnetic susceptibility
\bea
&& \chi(B=0) = - (2e w/c)^2\frac{\p^2 F}{\p Q^2}|_{Q=0} =\nonumber\\
&&  \frac{1}{2\pi v}(2e w /c)^2\frac{m_+m_-}{m_+^2 - m_-^2}\ln(m_+/|m_-|). 
\eea
changes its sign at the critical point $m_-=0$, being diamagnetic in the SC state.

At small wave vectors we have
\bea
&&\omega_-^2 \approx \\
&&(vk)^2\Big[1- \frac{(vQ)^2}{V_J\sqrt{V_c^2 +v^2Q^2}}\Big] + (V_J - \sqrt{V_c^2+Q^2})^2.\nonumber
\eea
Therefore the magnetic field shifts the critical point. According to \cite{shimshoni}, such a transition (at finite temperatures it becomes a crossover) from SC to PDW  state, instigated by magnetic field, has already been observed \cite{exp}. However, the shift of the transition point is not the only effect of the field. At 
\be
(vQ)^2 > V_J^2/2 + V_J\sqrt{V_J^2/4 + V_c^2},
\ee
the spectrum changes its form and instead of having one minimum at $k=0$ acquires two minima at 
\be
(k_{min}v)^2 = \frac{1}{(vQ)^2}\Big[(vQ)^4 - (vQ)^2V_J^2 - V_J^2V_c^2\Big],
\ee
The spectral gaps are $
\Delta = V_c\sqrt{1 - (V_J/vQ)^2}$. 
This leads to appearance of incommensurate oscillations in correlation functions.

Now imagine that $V_J,V_c$ are coordinate dependent and at some point one of the Majorana masses changes sign (since the Josephson tunneling $g_J$ strongly depends on the interchain distance, this can be easily achieved by changing the distance between the wires). Then there is  a zero energy Majorana mode bound to such a kink; its wave function is given by
\bea
\left(\begin{array}{c}
\eta_R^0(x)\\
\eta_L^0(x)
\end{array}
\right) = \frac{\gamma(x_c)}{{\cal N}} \re^{ \pm \int_{x_c}^x m_s(y) d y/v} \left(\begin{array}{c}
1\\
\mp 1
\end{array}
\right) + ...,  \label{zero}
\eea 
where $\gamma$ is the operator of ZEMM ($\gamma^2 =1$), ${\cal N}$ is the normalization factor and the dots stand for modes with non-zero energy. The sign in (\ref{zero}) depends on whether $m_s(x)$ is a kink or an anti-kink. ZEMM is robust with respect to changes in $K_-$, so far as both cosine operators in (\ref{charge2}) remain relevant  ($1/2< K_- <2$).  In a system with periodic boundary conditions there is always an anti-kink carrying another ZEMM $\gamma(x_{c2})$. Together two gamma-matrices create a two-dimensional irreducible representation, where creation and annihilation operators are defined as $\gamma(x_{c1}) \pm \ri\gamma(x_{c2})$. 

ZEMMs can be used to assist in  a coherent hopping ``teleportation") of massive Majorana fermions between different pairs of wires, as takes place in the Kitaev model. Consider two such pairs where the phase boundaries between the SC and the PDW phases are brought sufficiently close to each other, as on Fig.1. 
\begin{figure}[h]
\includegraphics[width=1.0\columnwidth]{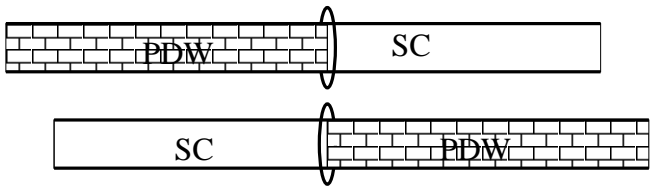}
\caption{This picture illustrates a coherent tunneling ("teleportation") scheme for massive Majorana fermions described in the text. Solid lines represent superconducting wires. Areas with different filling are PDW and SC phases. ZEMMs are located on the phase boundaries and are represented by the ellipses. }
 \label{fig-res1} 
\end{figure}
ZEMMs are involved in the smooth part of the operator 
\bea
&& (n_1 - n_2)_{smooth} = -\frac{1}{\sqrt{\pi K_-}}\p_x\Theta_- \rightarrow \nonumber\\
&& \ri \psi_0(x-x_c)\gamma(x_c)[\rho_R(x) \pm \rho_L(x)] , \label{tele}
\eea
where $\psi_0$ is the ZEMM wave function (\ref{zero})  and the sign depends on whether the SC phase is on the left or on the right.  The contribution of (\ref{tele}) to the repulsive interaction of pairs  is 
\bea
V = {\cal V}(r_{12})\gamma({\bf r}_1)\gamma({\bf r}_2)[\rho_R({\bf r}_1) - \rho_L({\bf r}_1)][\rho_R({\bf r}_2) + \rho_L({\bf r}_2)], \label{tele2}
\eea
(the signs correspond to the configuration depicted on Fig. 1), where ${\cal V}$ is the matrix element of the interaction and ${\bf r}_{1,2}$ are located on different pairs of wires. If the given pair of domain walls is sufficiently isolated from all others so that the given ZEMMs do not interact with other ones, then $N = \ri\gamma({\bf r}_1)\gamma({\bf r}_2) = \pm 1/2$ is an integral of motion and can be replaced by a $c$-number. Thus the four-fermion term (\ref{tele2}) becomes a two-fermion term describing a coherent hopping of massive $\rho$-fermions between the pairs of wires. 

It is interesting to note that the expression for the total superconducting current 
\bea
J = \frac{e\rho_s}{2mc}(\p_x\Phi_1 + \p_x\Phi_2) = \frac{e\sqrt {2\pi K_+}\rho_s}{2mc}\p_x\Phi_+,  \label{current}
\eea
does not depend on the masses. So, provided the stiffness $\rho_s$ does not change through the system, the current is the same in the SC and PDW phases! This reflects the fact that {\it in the absence of disorder}  the PDW  state supports a sliding density wave which is equivalent to phase coherent tunneling of superconducting pairs through the PDW region. However, with  disorder present, the phase $\Theta_+$ is likely to get pinned so that  the PDW regions become insulating. In the present context this was noted in \cite{shimshoni}.  It should be added however, that the Josephson coupling between two SC regions separated by a PDW one now depends on the amount of disorder in the PDW region and can, at least in principle, be made as strong as one likes by cleaning the sample.

From the point of view of quantum information applications such as considered in \cite{oppen} the most interesting situation is when boundaries between SC and PDW phases are few and far between. Then these modes do not interact and what remains of them is the Clifford algebra of  $\gamma_1,...\gamma_N$ operators ($N$ is the number of boundaries) which is equivalent to the algebra of Dirac $\gamma$-matrices. However, the situation when the zero modes overlap and form an impurity band is also interesting. This situation was considered in \cite{yulu},\cite{ShTs}, where it was found that the band contains one delocalized state at zero energy and gives rise to a singular  contribution to the specific heat:
\bea
C(T) \sim n_i\ln^{-3}(T_0/T), \label{CV}
\eea
where $n_i$ is the concentration of the boundaries and $T_0$ is a nonuniversal energy scale of the order of the impurity band width. As far as I can see, ZEMMs do not yield singular contributions to any other thermodynamic quantities or correlation functions except the thermal conductivity. The reason for this is that in all observables except the energy density ZEMMs appear in conjunction with operators for which correlation functions decay exponentially. 

 The disordered phase described in \cite{yulu},\cite{ShTs} is somewhat special. In the present context it is just a random sequence of SC and PDW phases with each the phases being essentially clean. In real wires there is another type of disorder, in the form of impurity potentials coupled to the particle density (\ref{dens}). As was rightly pointed out in \cite{shimshoni}, this disorder becomes effective in the PDW phase transforming it to an insulator.  I suggest, however, that the conventional disorder does not affect the ZEMM so that  the results of \cite{yulu},\cite{ShTs} including formula  (\ref{CV}) are robust. 
 Indeed, the most relevant part of the impurity contribution to the Hamiltonian comes from the coupling of the disorder potential to the $2\pi n_0$ Fourier harmonics of the density operator:
 \bea
&& \int U(x)\rho(x) \rd x  \rightarrow\\
 && \int \rd x \Big\{ \re^{\ri\sqrt{\pi/K_+}\Theta_+}\Big[ U_+(x)\s_1\mu_2 + \ri U_-(x)\mu_1\s_2\Big] + h.c.\Big\}, \nonumber
 \eea
where $U_+,U_-$ are Fourier envelopes of the disorder potential with wave vectors concentrated around $2\pi n_0$. In the SC phase $\la \s_a\mu_b\ra = 0$ and the disorder is not effective. In the PDW state one of the amplitudes acquires an average vacuum value and the most relevant contribution to the Hamiltonian (\ref{Gaussian}) becomes 
\bea
\delta H = A\int \rd x |U(x)|\cos[\sqrt{\pi/K_+}\Theta_+(x) - \alpha(x)], \label{deltaV}
\eea
where $A \sim (V_c^2-V_J^2)^{1/8}$ and $\alpha(x)$ is a random phase. This perturbation is relevant at $K_+ > 1/6$ and leads to a pinning of the phase field $\Theta_+$. However, the interaction predominantly affects the $\Theta_+$ field and not the Majorana modes. That the interaction between the latter and $\Theta_+$ is weak can be ascertained from the fact that the staggered density amplitude vanishes at the boundary between the SC and PDW phases where the zero modes are located. Using Eqs.(83,84) from \cite{yulu} we get ($x<0$ is in the PDW phase):
 \bea
 &&A(x) = \\
 &&  \la\s_1\mu_2\ra = (a_0^2m_+m_-)^{1/8} \exp\{- f[m_-(x-x_c)]\}, \nonumber\\
 && f(x) = \frac{1}{2}\theta(x) x + \frac{1}{8}\Big[K_0(|x|) +  K_{-1}(|x|) \Big], \nonumber
\eea
($a_0$ is the lattice cut-off), so that at the boundary where the Majorana zero mode  is located the amplitude of the staggered density  vanishes: $A(x) = \la\s_1\mu_2(x)\ra \sim |x-x_c|^{1/8}$.
\begin{figure}[h]
\includegraphics[width=1.0\columnwidth]{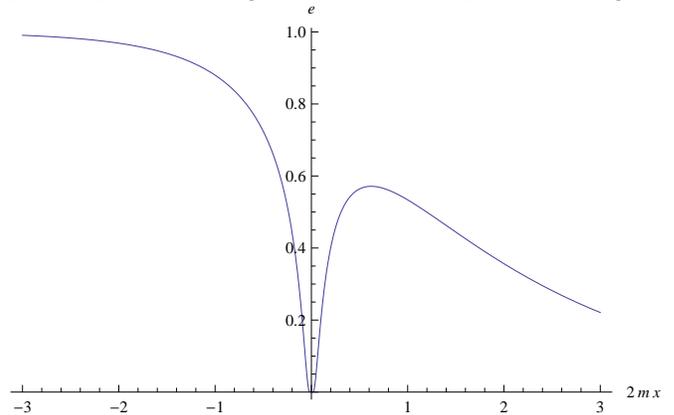}
\caption{ $e=A(x)/A(-\infty)$ as a function of $2m_-(x-x_c)$.}
 \label{fig-res2} 
\end{figure}

Thus I have demonstrated that in the model of coupled superconducting wires there is a parameter region where one can expect the existence of zero energy Majorana modes. I think that in the light of experiments \cite{exp} material realizations of such systems are  entirely realistic. Local changes of the model parameters required to create ZEMMs can be achieved either by local magnetic fields or by a change of the stripe width $w$. This may help to produce a material realization of the Kitaev model using conventional superconductors. 

 I am grateful to Efrat Shimshoni and Alexander Nersesyan  for interesting discussions. 
The work was  supported  by the Center for Emerging Superconductivity funded by the US DOE, Office of Science.

\end{document}